\documentclass[12pt]{article}
\usepackage{graphicx}
\usepackage{color}
\usepackage{cite}

\def\upstrut{\vrule height 2.5ex depth 0.0ex width 0pt}

\def \beq{\begin{equation}}
\def \eeq{\end{equation}}
\def\eqref#1{(\ref{#1})}
\def\bea{\begin{eqnarray}}
\def\eea{\end{eqnarray}}
\def\jpsi{\hbox{$J\kern-0.2em/\kern-0.1em\psi$}}

\def\URLtilde{\lower0.2em\hbox{$\tilde{\phantom{a}}$}}
\def\mycomm#1{\hfill\break\strut\kern-3em{\color{red}\tt ====> #1
\color{black}}\hfill\break}

\newcount\timecount
\newcount\hours \newcount\minutes  \newcount\temp \newcount\pmhours
\hours = \time
\divide\hours by 60
\temp = \hours
\multiply\temp by 60
\minutes = \time
\advance\minutes by -\temp
\def\hour{\the\hours}
\def\minute{\ifnum\minutes<10 0\the\minutes
\else\the\minutes\fi}
\def\clock{
\ifnum\hours=0 12:\minute\ AM
\else\ifnum\hours<12 \hour:\minute\ AM
\else\ifnum\hours=12 12:\minute\ PM
\else\ifnum\hours>12
\pmhours=\hours
\advance\pmhours by -12
\the\pmhours:\minute\ PM
\fi
\fi
\fi
\fi
}

\def\monthname{\relax\ifcase\month 0/\or January\or February\or
March\or April\or May\or June\or July\or August\or September\or
October\or November\or December\else\number\month/\fi}

\def\bold#1{\setbox0=\hbox{$#1$}     \kern-.025em\copy0\kern-\wd0
\kern.05em\copy0\kern-\wd0
\kern-.025em\raise.0433em\box0 }

\begin{document}
\setcounter{footnote}{1}
\rightline{EFI 15-20}
\rightline{TAUP 2997/15}
\rightline{arXiv:1506.06386}
\vskip1cm

\begin{center}
{\Large \bf New Exotic Meson and Baryon Resonances
\\
\upstrut
from Doubly-Heavy Hadronic Molecules}
\end{center}
\bigskip

\centerline{\bf Marek Karliner$^a$\footnote{{\tt marek@proton.tau.ac.il}}
 and Jonathan L. Rosner$^b$\footnote{{\tt rosner@hep.uchicago.edu}}}
\medskip

\centerline{$^a$ {\it School of Physics and Astronomy}}
\centerline{\it Raymond and Beverly Sackler Faculty of Exact Sciences}
\centerline{\it Tel Aviv University, Tel Aviv 69978, Israel}
\medskip

\centerline{$^b$ {\it Enrico Fermi Institute and Department of Physics}}
\centerline{\it University of Chicago, 5620 S. Ellis Avenue, Chicago, IL
60637, USA}
\bigskip
\strut

\begin{center}
ABSTRACT
\end{center}
\begin{quote}
We predict several new exotic doubly-heavy hadronic resonances, inferring from
the observed exotic bottomonium-like and charmonium-like narrow states 
\break
$X(3872)$, $Z_b(10610)$, $Z_b(10650)$, $Z_c(3900)$, and $Z_c(4020/4025)$.
We interpret the binding mechanism as mostly molecular-like isospin-exchange
attraction between two heavy-light mesons in a relative S-wave state.  We then
generalize it to other systems containing two heavy hadrons which can couple
through isospin exchange.  The new predicted states include resonances in
meson-meson, meson-baryon, baryon-baryon, and baryon-antibaryon channels.
These include those giving rise to final states involving a heavy quark
$Q=c,b$ and antiquark $\bar Q' = \bar c,\bar b$, namely
$D \bar D^*$,
$D^* \bar D^*$,
$D^* B^*$,
$\bar B B^*$,
$\bar B^* B^*$,
$\Sigma_c \bar D^*$,
$\Sigma_c B^*$,
$\Sigma_b \bar D^*$, 
$\Sigma_b B^*$,
$\Sigma_c \bar \Sigma_c$,
$\Sigma_c \bar \Lambda_c$,
$\Sigma_c \bar \Lambda_b$,
$\Sigma_b \bar \Sigma_b$,
$\Sigma_b \bar \Lambda_b$,
and
$\Sigma_b \bar \Lambda_c$,
as well as corresponding S-wave states giving rise to $Q Q'$ or $\bar Q \bar
Q'$.
\end{quote}

\smallskip

\leftline{PACS codes: 12.39.Hg, 12.39.Jh, 14.20.Pt, 14.40.Rt}
\bigskip



During the last few years there have been several experimental discoveries
of bottomonium-like and charmonium-like charged manifestly exotic narrow
isovector resonances $Z_b(10610)$, $Z_b(10650)$ \cite{Abe:2007tk,%
Karliner:2008rc,Belle:2011aa,Adachi:2012cx,Krokovny:2013mgx,Garmash:2014dhx,%
Krokovny:Jinan}, $Z_c(3900)$ \cite{Ablikim:2013mio,Liu:2013dau,Xiao:2013iha,%
Ablikim:2013xfr,BESIII:2015kha}, and $Z_c(4020/4025)$ \cite{Ablikim:2013emm,%
Ablikim:2013wzq,Chilikin:2014bkk,Ablikim:2014dxl,Guo:Jinan}.
All four resonances lie very close to two heavy meson thresholds:
$\bar B B^*$, $\bar B^* B^*$, $\bar D D^*$ and $\bar D^* D^*$,
respectively. The discoveries of these states were preceded by the 
observation of the now well-established $X(3872)$ \cite{X3872}
extremely narrow resonance right at the $\bar D D^*$ threshold.  
Conspicuously absent from this list 
are resonances at the $\bar D D$ and $\bar B B$ thresholds.

In all the states where $J^P$ has been unambiguously measured it is $1^+$.
The charged states decay into heavy quarkonia (e.g., \jpsi, $\psi^\prime$,
$\Upsilon$, or $h_b$) and a charged pion.\footnote{Some of the states also
decay to two heavy mesons, typically with much larger branching ratios, despite
much smaller phase space.} So they are manifestly exotic and 
their minimal quark content is $\bar Q Q q \bar q$, i.e., that of a
tetraquark, where $Q=c,b$ and $q=u,d$.

Yet, despite large phase space (hundreds of MeV) for decay into $\bar Q Q$ and 
pion(s), these resonances have narrow widths, indicating a very small overlap
of their wave functions with the corresponding quarkonia.  This provides strong
circumstantial evidence in favor of molecular interpretation, namely that
rather than containing all four quarks in a single confinement volume, the
resonances are loosely bound S-wave states of heavy-light mesons,
\hbox{$Q\bar q$-$\bar Q q$.}

Such ``molecular" states, $\bar D D^*$, etc., were introduced
in Refs.~\cite{Voloshin:1976ap} and \cite{DeRujula:1976qd}.  
They were later
extensively discussed 
\cite{deusons,Tornqvist:2004qy,Thomas:2008ja,Suzuki:2005ha,Fleming:2007rp}
in analogy with the deuteron which binds via exchange of pions and
other light mesons.

A crucial element of the binding mechanism proposed in \cite{deusons} is that
pions are expected to play a major role in generating the attractive potential.%
\footnote{There must also be a shorter range repulsive force to stabilize the
interaction.} From 
today's perspective one may generalize this to exchange of light 
quarks in their lowest-mass configuration, i.e., a pseudoscalar carrying one
unit of isospin. Such a binding mechanism immediately explains the 
conspicuous absence of $\bar D D$ and $\bar B B$ among the observed 
resonances. A resonance in a channel containing two heavy pseudoscalar
mesons cannot form through exchange of a pseudoscalar pion, because
such an exchange would require a three-pseudoscalar vertex, e.g.,
$DD\pi$, which is forbidden in QCD by parity conservation.

On the other hand, $\bar D D^*$ and $\bar D^* D^*$ (and their bottomonium
counterparts) can bind through pion exchange. In the $\bar D D^*$ case
$\bar D$ emits a pion and turns into $\bar D^*$, while $D^*$ absorbs a pion
and turns into $D$, so $\bar D D^* \to \bar D^* D$, etc. The physical state
is $(\bar D D^* + \bar D^* D)/\sqrt{2}$, so it turns into itself.

In the $\bar D^* D^*$ case, a $D^*$ can emit a pion and remain a $D^*$.
$D^*$ has negative parity, so the emitted pion must be emitted in a
$P$-wave. The orbital angular momentum can couple with $S=1$ from 
$D^*$ spin to give a total $J=1$. The same argument applies to 
$\bar D^*$, so $\bar D^* D^*$ turns into itself after pion emission.

Thus the conditions for existence of the resonance are
\begin{enumerate}
\item[(a)]
The state contains two heavy hadrons. They have to be heavy, as the repulsive
kinetic energy is inversely proportional to the reduced mass (see, e.g.,
\cite{Ericson:1993wy}).  (For a more recent discussion see \cite{Li:2014gra}.)
\item[(b)]
The two hadrons carry isospin, so that they can couple to pions.
Channels like $\Sigma_c \bar \Lambda_c$, in which one of the particles has
zero isospin, can exchange a pion to become the equal-mass channel $\Lambda_c
\bar \Sigma_c$.
\item[(c)]
The spin and parity of the two hadrons have to be such that they can bind
through single pion exchange.
\item[(d)] The hadrons making up the molecule have to be sufficiently narrow,
as the molecule's width cannot be smaller than the sum of its constituents'
widths \cite{Hanhart:2010wh,Filin:2010se,Guo:2011dd,MK:unpublished}.

\end{enumerate}

\noindent
Methods have been proposed \cite{Cleven:2015era} to distinguish molecular
states of two heavy mesons from alternative models.  Our discussion is confined
to molecular states, but includes meson-baryon, baryon-baryon, and
baryon-antibaryon channels.  Exotic baryon-antibaryon resonances were proposed
earlier \cite{Rosner:1968si}, but without the additional binding conferred by
a heavy quark-antiquark pair they would probably be too broad to detect.  The
binding mechanism can apply to two heavy baryons leading to a prediction of a
doubly heavy $\Sigma_b^+\Sigma_b^-$ dibaryon \cite{Karliner:2011yb,Karliner:%
2012pc,Karliner:2013dqa}.  We emphasize that {\em the pion-exchange binding
mechanism can in principle apply to any two heavy hadrons which carry isospin
and satisfy condition (c) above, be they mesons or baryons.}

For pion exchange between states 1 and 2 with isospins $I_{1,2}$ and spins
$S_{1,2}$, the effective potential is proportional to \cite{deusons}
\beq \label{eqn:int}
V \sim \pm (I_1 \cdot I_2) (S_1 \cdot S_2)~{\rm for}~(q q,q \bar q)~
{\rm interactions}~,
\eeq
where $q$ or $\bar q$ stands for the light quark(s) or antiquark(s) in
hadrons 1 and 2, as long as the total spins $S_i$ are correlated with the
direction of the light-quark spins.  (This is true for $D^*,~B^*,~\Sigma_c$,
and $\Sigma_b$.)

The new states we are discussing are expected to be narrow, just like the $Z_c$
and $Z_b$ states.  They are either below threshold or slightly above threshold
with regard to the two body channels in which pion exchange occurs.  So there
will be little or no phase space for decay into such channels.  On the other
hand, even though they will have plenty of phase space for decay into
quarkonium and states made from light quarks, their wave functions will have
small overlap with such final states.  This is because they are loosely bound
and therefore in the initial wave function the heavy quarks spend most of
their time far from each other.  Resonances are possible also in states with
higher isospin, but their masses are expected to be higher, and their widths
are expected to be larger, e.g.,
\bea
M(Z_c(4020/4025)) &>& M(X(3872))~,
\nonumber \\
\\
\Gamma(Z_c(4020/4025)) &\gg& \Gamma(X(3872))~,
\nonumber
\label{mass.width.vs.isospin.eq}
\eea
the latter because of the larger phase space for the ``fall-apart'' mode into
two heavy mesons.

A quick inspection leads to the following most likely candidates containing
a heavy quark $Q = c$ or $b$ and a heavy antiquark $\bar Q' = \bar c$ 
or $\bar b$:
$D \bar D^*$, $D^* \bar D^*$,
$D^* B^*$,
$\bar B B^*$,
$\bar B^* B^*$,
$\Sigma_c \bar D^*$,
$\Sigma_c B^*$,
$\Sigma_b \bar D^*$, 
$\Sigma_b B^*$, 
$\Sigma_c \bar \Sigma_c$,
$\Sigma_c \bar \Lambda_c$,
$\Sigma_c \bar \Lambda_b$,
$\Sigma_b \bar \Sigma_b$,
$\Sigma_b \bar \Lambda_b$,
and
$\Sigma_b \bar \Lambda_c$.
As noted above, these are the states whose heavy-quark content is $c \bar
c,~b \bar b,~b \bar c$, or $c \bar b$.  The first two types of states can decay
strongly to charmonium or bottomonium plus pion(s), while the latter two
involve a $B_c^\pm$ in the final state.  (This could provide a distinctive
signature at the LHC \cite{Li:2014gra}.)  There will also be corresponding
states (such as the $\Sigma_b \Sigma_b$ dibaryon proposed in Refs.\
\cite{Karliner:2011yb,Karliner:2012pc,Karliner:2013dqa}) whose heavy-quark
content is $Q Q'$ or $\bar Q \bar Q'$.  The
thresholds and some sample decay modes for the states with heavy-quark content
$Q \bar Q'$ are displayed in Table \ref{tab:thr}.

We have listed in Table \ref{tab:thr} channels which can undergo transitions
either to themselves or to equal-mass channels via pion exchange.  The channels
$\Sigma_c \bar \Lambda_b$ and $\Lambda_c \bar \Sigma_b$ are the sole
exception, which we have listed for the purpose of discussion.  Pion
exchange permits the transitions $\Sigma_c \bar \Lambda_b \leftrightarrow
\Lambda_c \bar \Sigma_b$ and $\Sigma_b \bar \Lambda_c \leftrightarrow
\Lambda_b \bar \Sigma_c$, channels whose thresholds differ by 27.6 MeV from
one another.  Another pair (not listed in the table) whose thresholds differ
by only 26 MeV are $\Sigma_c \bar D$ (threshold 4321 MeV) and $\Lambda_c \bar
D^*$ (threshold 4295 MeV).  It will be interesting to see if such nearby
thresholds have any role in fostering pion exchange.

\begin{table}[t]
\caption{\small\baselineskip12pt
Thresholds for molecular states consisting of a hadron with a heavy
quark $Q=c$ or $b$ and an antiquark $\bar Q' = \bar c$ or $\bar b$.  Similar
thresholds hold for states with $Q Q'$ or $\bar
Q \bar Q'$.  For non-self-conjugate cases, charge-conjugate channels are also
implied.  Here $q$ represents a light quark $u$ or $d$.  Only states which can
undergo transitions to equal-mass channels via pion exchange are shown.
Isospin violation in hadron masses is ignored. Charge-conjugate baryonic states
have opposite parity.
\label{tab:thr}}
\begin{center}
\begin{tabular}{c c c c c c} \hline \hline
Channel & Minimum &  Minimal quark  & Threshold & S-wave & Example of \\
        & isospin & content$^{a,b}$ & (MeV)$^c$ &  $J^P$ & decay mode \\ \hline
$D \bar D^*$ & 0  & $c \bar c q \bar q$ & 3875.8 & $1^+$& $\jpsi\, \pi \pi$ \\
$D^* \bar D^*$ & 0 & $c \bar c q \bar q$ & 4017.2 & $0^+,1^+,2^+$ &
 $\jpsi\, \pi \pi$ \\
$D^* B^*$ & 0 & $c \bar b q \bar q$ & 7333.8 & $0^+,1^+,2^+$ & 
 $B_c^+ \omega$ \\
$\bar B B^*$ & 0 & $b \bar b q \bar q$ & 10604.6 & $1^+$ &
 $\Upsilon(nS) \omega $ \\
$\bar B^* B^*$ & 0 & $b \bar b q \bar q$ & 10650.4 & $0^+,1^+,2^+$ &
 $\Upsilon(nS) \omega $ \\
$\Sigma_c \bar D^*$ & 1/2 & $c \bar c q q q'$ & 4462.4 & $1/2^-,3/2^-$ &
 $\jpsi\, p$ \\
$\Sigma_c B^*$ & 1/2 & $c \bar b q q q'$ & 7779.5 & $1/2^-,3/2^-$ &
 $B_c^+ p$ \\
$\Sigma_b \bar D^*$ & 1/2 & $b \bar c q q q'$ & 7823.0 & $1/2^-,3/2^-$ &
 $B_c^- p$ \\
$\Sigma_b B^*$ & 1/2 & $b \bar b q q q'$ & 11139.6 & $1/2^-,3/2^-$ &
 $\Upsilon(nS) p$ \\
$\Sigma_c \bar \Lambda_c$ & 1 & $c \bar c q q' \bar u \bar d$ & 4740.3 &
 $0^-,1^-$ & $\jpsi\, \pi$ \\
$\Sigma_c \bar \Sigma_c$ & 0 & $c \bar c q q' \bar q \bar q'$ & 4907.6 &
 $0^-,1^-$ & $\jpsi \,\pi \pi$ \\
$\Sigma_c \bar \Lambda_b$ & 1 & $c \bar b q q' \bar u \bar d$ & 8073.3$^d$ &
 $0^-,1^-$ & $B_c^+ \pi$ \\
$\Sigma_b \bar \Lambda_c$ & 1 & $b \bar c q q' \bar u \bar d$ & 8100.9$^d$ &
 $0^-,1^-$ & $B_c^- \pi$ \\
$\Sigma_b \bar \Lambda_b$ & 1 & $b \bar b q q' \bar u \bar d$ & 11433.9 &
 $0^-,1^-$ & $\Upsilon(nS) \pi$ \\
$\Sigma_b \bar \Sigma_b$ & 0 & $b \bar b q q' \bar q \bar q'$ & 11628.8 &
 $0^-,1^-$ & $\Upsilon(nS) \pi \pi$ \\
\hline \hline
\end{tabular}
\end{center}
\leftline{$^a$Ignoring annihilation of quarks.  $^b$Plus other charge states
when $I \ne 0$.}
\leftline{$^c$Based on isospin-averaged masses.  $^d$Thresholds differ by 27.6
MeV.}
\end{table}

A detailed analysis such as that of Refs.~\cite{deusons,Tornqvist:2004qy,%
Thomas:2008ja,Suzuki:2005ha,Fleming:2007rp},
is needed to determine whether pion exchange is sufficient to
bind two hadrons in each of the channels listed in Table \ref{tab:thr} or the
corresponding states with $Q Q'$ or $\bar Q \bar Q'$.
For the case of $D \bar D^*$ and $B \bar B^*$ we include here a detailed
example of mixing between channels which have equal mass in the isospin
symmetry limit.  For the former, the four channels are
\beq
[D^0 \bar D^{*0},~D^{*0} \bar D^0,~D^+ D^{*-},~D^{*+} D^-]~.
\eeq

In this basis pion exchange leads to a potential proportional to the
matrix
\beq
V \sim \left[ \begin{array}{r r r r} 0 & -1 &  0 & -2 \\
                                    -1 &  0 & -2 &  0 \\
                                     0 & -2 &  0 & -1 \\
                                    -2 &  0 & -1 &  0 \end{array} \right]
\eeq
The eigenvalues and eigenvectors are
\bea
 1&:&~~[1,~~~~1,~-\!1,~-\!1]~~~C = +,~I = 1~,\\
-1&:&~~[1,~-\!1,~-\!1,~~~~1]~~~C = -,~I = 1~,\\
 3&:&~~[1,~-\!1,~~~~1,~-\!1]~~~C = -,~I = 0~.\\
-3&:&~~[1,~~~~1,~~~~1,~~~~1]~~~C = +,~I = 0~.
\eea
The last corresponds to the most deeply bound state $X(3872)$.  The state with
eigenvalue $-1$, negative $C$, and $I=1$ can be identified with the $Z_c(3900)$.

In this particular case the mixing matrix is $4 \times 4$.  In the real world
the mass difference between $D^0 \bar D^{*0}$ and $D^+ D^{*-}$ is much larger
than the binding energy, so the physical $X(3872)$ is reduced to a two-channel
mixture of $D^0 \bar D^{*0}$ and $D^{*0} \bar D^0$.  In the $B \bar B^*$ case
isospin breaking is much smaller and the binding is expected to be stronger,
because of the larger reduced mass, leading to smaller kinetic energy.
Therefore in the case of $X_b$, the bottomonium analogue of $X(3872)$, one
expects a full four-channel mixing.

Analogous mixing is expected in other S-wave meson-meson, meson-baryon, and
baryon-(anti)baryon channels.  The channels $D^{*0} \bar D^{*0}$
and $D^{*+} D^{*-}$ constitute three separate two-channel problems for
states of total spin $J=0$, 1, and 2, with eigenchannels corresponding to
isospin $I=0$ and 1, while the channels $\Sigma_c \bar D^*$ constitute two
separate two-channel problems for $J=1/2$ and 3/2 with eigenchannels
corresponding to $I=1/2$ and $3/2$.

The expression (\ref{eqn:int})
predicts the most attractive $D^* \bar D^*$ channel to have $I=S=0$.
The only state discovered so far near that threshold is $Z_c(4020/4025)$ with
$I=1$. In addition to $Z_c(4020/4025)$ there could be another state near $D^*
\bar D^*$ threshold, with lower mass and $I=0$.

The next threshold above $D^* \bar D^*$ in Table \ref{tab:thr} is that of
$\Sigma_c \bar D^*$, at 4462 MeV.  Application of Eq.\ (\ref{eqn:int}) with a
+ sign for $qq$ interaction predicts two lowest levels with the same binding
energy: $S=1/2,~I=3/2$ and $S=3/2,~I=1/2$.  The $J/\psi p$ mode listed in the
table can only come from the latter (spin-3/2) state. 

As little is known about pion couplings to most of the states in Table
\ref{tab:thr}, and as exchanges other than pions and configurations other than
S waves (as discussed, e.g., in Ref.\ \cite{deusons}) may play a role, it is
too early to calculate the binding in most cases.  Our purpose here is to call
attention to some interesting possible thresholds whose effects could show up
in final states consisting of heavy quarkonium plus light-quark mesons or
baryons.  Such final states are accessible in several current experiments.

{\it Notes added:} We thank X. Liu for informing us of an earlier calculation
\cite{Yang:2011wz} of binding between a charmed baryon and anticharmed meson,
obtaining --- as we do --- no binding between $\Lambda_c$ and $\bar D^{(*)}$
but binding between $\Sigma_c$ and $\bar D^*$ in all four spin--isospin
channels, as well as 
--- unlike us --- between $\Sigma_c$ and $\bar D$ with $I=3/2$ and $J=1/2$.

The LHCb Collaboration has just posted results \cite{LHCb} on a new narrow
exotic resonance $P_c(4450)$
in the $J/\psi p$ channel, with a mass of $4449.8\pm1.7\pm2.5$ MeV, a width
of $39\pm5\pm8$ MeV, and statistical significance $12 \sigma$.  Its mass and
spin are consistent with the $\Sigma_c \bar D^*$
$I=1/2,~S=3/2$ resonance that we predict based on Table I.

In the same paper LHCb reports discovering another, lighter and wider
Breit-Wigner structure $P_c(4380)$, also in the $J/\psi p$ channel, with a mass
of $4380 \pm 8\pm 29$ MeV and a width of $205\pm 18\pm 86$ MeV.
This structure is {\em not} predicted by our approach.
At this point it isn't clear if $P_c(4380)$ is a regular resonance, because
of the unusual shape of its Argand plot in Fig.~9(b) of 
Ref.~\cite{LHCb}, in contradistinction with the pristine plot
for $P_c(4450)$ in Fig.~9(a), 
though this could just be due to smaller statistics.
If it is not a {\em bona fide} resonance,
it is possible that $P_c(4380)$ results from 
the vicinity of the threshold, e.g., along the lines discussed in 
Ref.~\cite{Voloshin:2015ypa}.

If $P_c(4380)$ does turn out to be a genuine resonance after all, it is 
is very unlikely to be of molecular nature. This is because of the deep
binding, about 80 MeV below the $\Sigma_c \bar D^*$ threshold and the rather
large width. Instead, it would likely be some kind of a ``genuine"
$P$-wave pentaquark, for which both the large binding and the large width 
are much more natural. $P$-wave would then be essential in order for 
$P_c(4380)$ and $P_c(4450)$ to have opposite parities, as reported by LHCb.
The true nature of $P_c(4380)$ is an intriguing issue which is an outstanding
challenge for future experiments.  In particular, three recent papers
\cite{Wang:2015jsa,Kubarovsky:2015aaa,Karliner:2015} propose photoproduction
off a proton target as a test for the resonant nature of the enhancements
at 4380 and 4450 MeV.

We thank F. Close, C. Hanhart, X. Liu, V. Yu. Petrov, and C. E. Thomas for
helpful communications.  The work of J.L.R. was supported in part by the U.S.
Department of Energy, Division of High Energy Physics, Grant No.\
DE-FG02-13ER41958, and performed in part at the Aspen Center for Physics, 
which is supported by National Science Foundation grant PHY-1066293.

\end{document}